\documentclass[pre,aps,floats,superscriptaddress,floatfix,twocolumn]{revtex4}
\usepackage{amssymb,amsmath}
\usepackage{amsmath,amssymb}
\usepackage{graphicx}
\usepackage{psfrag}
\usepackage{color}
\usepackage{dcolumn}
\usepackage{bm}
\usepackage[normalem]{ulem}

\def\beq{\begin{equation}}
\def\eeq{\end{equation}}
\def\bea{\begin{eqnarray}}
\def\eea{\end{eqnarray}}
\begin{document}
\title{Asymmetric exclusion processes with fixed resources: Reservoir crowding and steady states}
\author{Astik Haldar}\email{astik.haldar@gmail.com}
\affiliation{Condensed Matter Physics Division, Saha Institute of
Nuclear Physics, HBNI, Calcutta 700064, West Bengal, India}
\author{Parna Roy}\email{parna.roy14@gmail.com}
\affiliation{Shahid Matangini Hazra Government College for Women, Purba Medinipore 721649, West Bengal, India}
\author{Abhik Basu}\email{abhik.123@gmail.com, abhik.basu@saha.ac.in}
\affiliation{Condensed Matter Physics Division, Saha Institute of
Nuclear Physics, HBNI, Calcutta 700064, West Bengal, India}
\begin{abstract}
 We study the nonequilibrium steady states of an asymmetric exclusion process (TASEP) coupled to a reservoir of unlimited capacity. We elucidate how 
 the steady states are controlled by the  interplay between the reservoir population that dynamically controls both the entry and exit rates of the 
 TASEP, and the total particle number in the system. The TASEP can be in the low density, high density, maximal current and shock phases. 
 We show that such a TASEP is  different from an open TASEP for {\em all} values of available resources: here, the TASEP can support only localised domain walls for any (finite) amount of resources
 as opposed to delocalised domain walls in open TASEPs. Furthermore, in the limit of infinite resources, the TASEP can be found in its high density phase only for any finite values of the control parameters, in contrast to an open TASEP.
\end{abstract}
\maketitle

\section{Introduction}

Totally asymmetric simple exclusion process (TASEP) was originally introduced as a conceptual model for describing quasi one-dimensional (1D) motion of molecular motors along microtubules in eukaryotic cells~\cite{macdonald}. 
Later on it emerged as a paradigmatic model for boundary-induced nonequilibrium phase transitions in 1D~\cite{krug}. The underlying  microscopic dynamics 
of TASEP violates the condition of detailed balance, and as a result, the ensuing steady state of TASEP is a genuine nonequilibrium steady state 
(NESS)~\cite{tasep-rev}, with a phase diagram controlled by the boundary rates of particle injection to and extraction from the TASEP, i.e., essentially the interactions
of the system with its environment. Generalisations and variants of TASEPs have more recently been studied as minimal models for  wide classes of physical phenomena, including traffic flow~\cite{traffic} and biological transport~\cite{bio-trans}.

Unlike a TASEP with open boundary conditions, TASEPs in closed geometries strictly conserve the total particle number. An interesting class of models belonging to this category consists of one or more TASEPs connected to a reservoir with a global particle number conservation~\cite{zia1,zia2,zia3,greulich,brackley}.  These models were introduced to describe the limited availability of resources required for a given physical
or biological process, e.g., finite availability of ribosomes in messenger RNA translocation for protein synthesis in eukaryotic cells, or a finite number of vehicles in a closed network of roads with a toll plaza or a drive-in (any place where several vehicles would wait at any time for some specific purposes, that may serve as a ``pool'' or ``reservoir'' of vehicles). Various possibilities including multiple TASEP lanes connected to a single point reservoir and the presence of fuel carriers in addition to the particles have been investigated in these studies. In these models, the primary effect of the finiteness of the available resources is that the effective entry rate of the particles to the TASEP, e.g., the actual protein synthesis taking place, sensitively depends upon the available resources. As the resources go up (reduce), protein synthesis rates increase (decrease). From the standpoint of nonequilibrium statistical mechanics, these models serve as minimal models for ``boundary induced 
nonequilibrium phase transitions with a conservation law''. Principal findings from these studies include 
modification of the phase diagram of an open TASEP and emergence of a shock phase characterised by localised domain walls (LDW). Unsurprisingly, strict particle number conservation plays a pivotal role in determining the NESS in these models. It is generally expected that in the limit of diverging amount of resources, the effects of particle number conservation are expected to be less important, and consequently, the NESS of the TASEP should largely resemble the one with open boundary conditions with delocalisation of the domain walls with equal effective entry and exit rates~\cite{parna-anjan}.

In this work, we explore a mechanism for avoidance of a crowded reservoir by the particles or agents, i.e.,the particles are {\em more inclined} to leave the reservoir, but {\em more inhibited} to return to it, if the latter gets more crowded. We show that this ensures that even in the limit of infinite resources, the TASEP connected to a reservoir remains 
qualitatively different from a TASEP with open boundary conditions. We call this ``reservoir crowding effect'', and model this in terms 
of ``effective entry and exit rates'', both of which depend upon the instantaneous reservoir occupation.  We  ask how this can affect the NESS of the 
TASEP connected to it.  It is reasonable to expect that a crowded reservoir (i.e., with ``high'' reservoir occupation) not only facilitate entry of 
particles into the TASEP, but may also hinders particles leaving TASEP as well.  This can potentially lead to a very low current in the TASEP in the steady 
state in the limit of high reservoir occupation. To this end we introduce a simple minimal model consisting of a reservoir without any spatial extent and 
a single TASEP lane that is connected to the reservoir at both its ends, and study this model systematically. Our principal results are as follows. Although this model in its NESS admits four distinct phases - low density (LD), high density (HD), maximal current (MC) and shock (SP) phases - just like a TASEP with open boundaries, there are significant differences: (i) In general, the phase diagram of this model is very different from that of an open TASEP as parametrised by the model parameters; by 
tuning the model parameters that define the particle inflow to and outflow from TASEP, the TASEP can preferentially be  populated or depopulated. More specifically, (ii) the well-known first order transition between the LD and HD phases in an open TASEP is replaced by 
{\em two} second order transitions, one between the LD and SP phases, 
and another between the SP and HD phases. The system shows generic LDWs in the SP phase. Unexpectedly and in contrast to the results in Ref.~\cite{parna-anjan}, 
the LDWs remain pinned for {\em any} particle number or any resources in the system, and show no tendency to delocalise even in the limit of very large resources for which the effects of the particle number conservation is na\"ively expected to be unimportant.   (iii) In the limit of infinite resources, the TASEP can only be in its HD phase, in stark contrast to an open TASEP, or to other existing models of TASEPs connected to a reservoir. This makes the present model fundamentally different.

The rest of the article is organised as follows. In Sec.~\ref{state}, we heuristically argue the nature of the steady states. Then in Sec.~\ref{phase} we show a series of phase diagrams in the $\alpha-\beta$ plane for various values of $\mu$, which highlight the sensitive dependences of the phase boundaries on $\mu$. Next, in Sec.~\ref{model}, we have introduced and defined the model. Then in Sec.~\ref{mft}, mean field theory (MFT) analysis of our model along with the phase diagrams and the steady state density profiles, complemented by extensive Monte-Carlo simulation (MCS) studies, are presented. The in Sec.~\ref{phase-trns}, we discuss the nature of the phase transitions in the models. Then in Sec.~\ref{dw} we analyse the nature of the domain walls. We end by summarising 
our results in Sec.~\ref{sum}.

\section{Model}\label{model}
\begin{figure}
 \includegraphics[width=7cm]{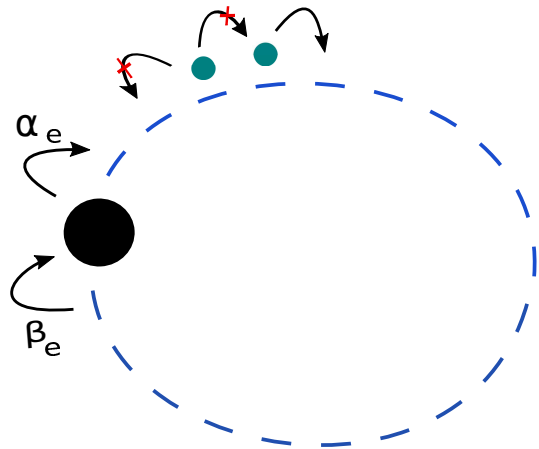}
 \caption{Schematic model diagram: the small filled circle $R$ is the point reservoir of infinite capacity; the broken line $T$  is the TASEP lane with $L$ sites connected to $R$ at both 
 the ends, with $\alpha_e$ and $\beta_e$ being the effective entry and exit rates of $T$ (see text).}\label{schem}
\end{figure}
The model consists of a single TASEP lane $T$ connected at both its ends to a reservoir $R$. The particles from $R$ enter $T$ through its entry end, 
hop unidirectionally along $T$ subject to exclusion and eventually leave $T$ at its exit end and enter back into $R$. Due to the closed geometry 
of the system, the dynamics clearly conserves the total particle number $N_0$. The reservoir $R$ is a point reservoir, without any spatial extent or internal dynamics, and can accommodate any number of particles without any upper limit. 

TASEP lane $T$ has $L$ sites, which are labelled by $i$; $i\in [1,L]$ with $i=1$ and $i=L$ being at the entry and exit sides, respectively. 
The entry and exit rates of $T$ are parametrised by $\alpha$ and $\beta$, respectively, which can take any positive values without restrictions. The actual entry and exit rates are dynamically controlled, and are given by
\begin{equation}
 \alpha_e=\alpha f(N),\;\beta_e=\beta g(N),
\end{equation}
where $N$ is the instantaneous occupation of $R$. We define filling factor $\mu = N_0/L$, which describes the population of the whole model ($T$ and $R$ combined) relative to the size $L$ of $T$; thus $0\leq \mu \leq \infty$. Our model, therefore, is a three-parameter model - the NESS are parametrised by $\alpha,\,\beta$ and $\mu$. Rate functions $f(N)$ and $g(N)$ control the actual entry and exit of particles to/from $T$. 
Since we are considering a situation where enhanced particle content in $R$ leads to a greater inflow of particles into $T$ and hinders 
outflow of particles from $T$ to $R$, we assume $f(N)$ and $g(N)$ to be monotonically increasing and decreasing functions of $N$. In order to 
reduce the parameter space further, we set $f(N)+g(N)=1$ together with the specific choice 
\begin{equation}
f(N)=N/N_0,\label{func}
\end{equation}
for simplicity.  The choice (\ref{func}) implies that the actual entry rate is proportional to the reservoir occupation $N$.
Thus $f(N)$ rises monotonically with $N$ with $f(0)=0$ and $f(N=N_0)=1$. The choice (\ref{func}), though similar, but is different quantitatively from the existing models for TASEPs with finite resources~\cite{zia1,zia2,zia3,greulich}; this suffices for our purposes and are easily amenable to analytical MFT treatments. 
Further, the choice (\ref{func}) implies 
\begin{equation}
g(N)=1-f(N)=1-N/N_0. \label{g-func}
\end{equation}
Thus, $g(N)$ is monotonically decreasing with the reservoir occulation $N$ with $g(0)=1$ and $g(N=N_0)=0$. We thus see that as $N$ rises, $f(N)$ rises but $g(N)$ reduces. This implies that as $N$ increases, more particles  try to enter into the TASEP lane and less number of particles would be able to leave it. In particular, for large $N_0$, when most of the particles are in $R$, $\alpha_e$ approaches $\alpha$ and $\beta_e$ becomes very small.  This feature is in contrast to the existing models for TASEPs with finite resources where in the large resources limit, the effective entry  rates become constants identical to the parameters that characterise the effective entry rates~\cite{zia1,zia2,zia3,brackley,greulich}; the exit rates are usually taken to be a given model parameter independent of any reservoir occupation (just as in open TASEPs). Our choices (\ref{func}) and (\ref{g-func}) are simple and minimal choices that are easily analytically tractable, and suffice to study the crowding effect 
introduced above.
Since 
the only condition on $f(N)$ and $g(N)$ is that these functions must be non-negative, $N$ can go up to $N_0$. Since $N_0$, the total number of particles, can even be 
infinity, the reservoir can contain an unlimited number of particles. 

\section{Steady state densities}\label{state}

Let $n_i$ be the occupation at site $i$ and $J$ be the corresponding current. The latter is a constant in the steady states. Below we outline 
the mean-field theory (MFT) and use it to obtain the phases  and phase diagram in the $\alpha-\beta$ plane, parametrised by $\mu$, supported 
by our extensive MCS studies. Before we embark on our MFT analysis, we already note that for small $\mu$, $\alpha_e$ should be small, where as $\beta_e$ should be large. Thus $T$ should not be in its HD phase. In fact, if $\mu < 1/2$, there are not enough particles in the system to keep $T$ in its HD or MC phases. Thus, for $\mu <1/2$, $T$ is {\em always} in its LD or SP phases independent of $\alpha$ and $\beta$. In contrast, for very large $\mu$, $\alpha_e \gg \beta_e$, and $T$ is expected to be in the HD phase only. For intermediate values of $\mu$, transition to the MC phase may be observed. We will see below that our detailed MFT analysis corroborates this intuitive physical picture.

\subsection{Phase diagrams}\label{phase}

We first summarise our results in terms of a series of phase diagrams in 
the $\alpha-\beta$ plane parametrised by $\mu=0.6,\,1,\,2,\,1000,\,10000$; see Fig.~\ref{phase1}, Fig.~\ref{phase4}, Fig.~\ref{phase2}, Fig.~\ref{phase3} and Fig.~\ref{phase5} below. As revealed by these phase diagrams, the phases and the phase boundaries in the $\alpha-\beta$ plane sensitively depend upon the value of $\mu$.

\begin{figure}
 \includegraphics[width=9cm]{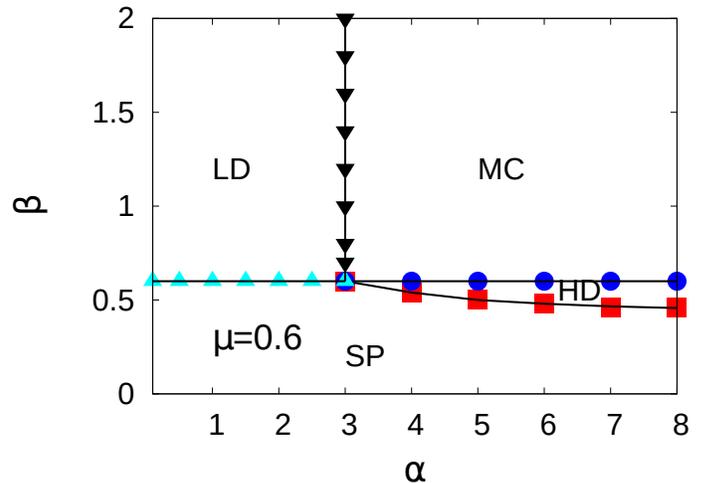}
 \caption{Phase diagram for $\mu=0.6$ in the $\alpha$ and $\beta$ plane. Continuous lines represent the MFT predictions (see text); the discrete points are the corresponding MCS results, which agree well with the MFT predictions. }\label{phase1}
\end{figure}
\begin{figure}
 \includegraphics[width=9cm]{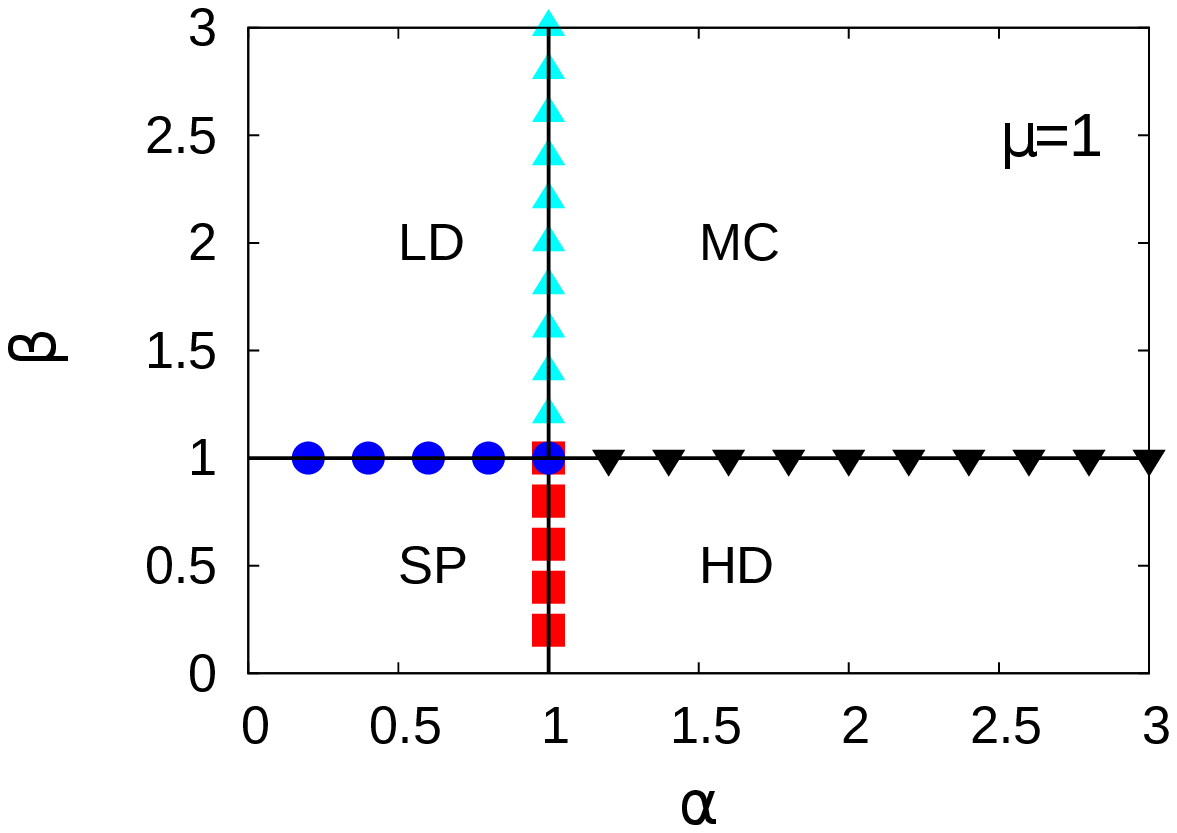}
 \caption{Phase diagram for $\mu=1$ in the $\alpha$ and $\beta$ plane. Continuous lines represent the MFT predictions (see text); the discrete points are the corresponding MCS results, which agree well with the MFT predictions.  }\label{phase4}
\end{figure}
\begin{figure}
 \includegraphics[width=9cm]{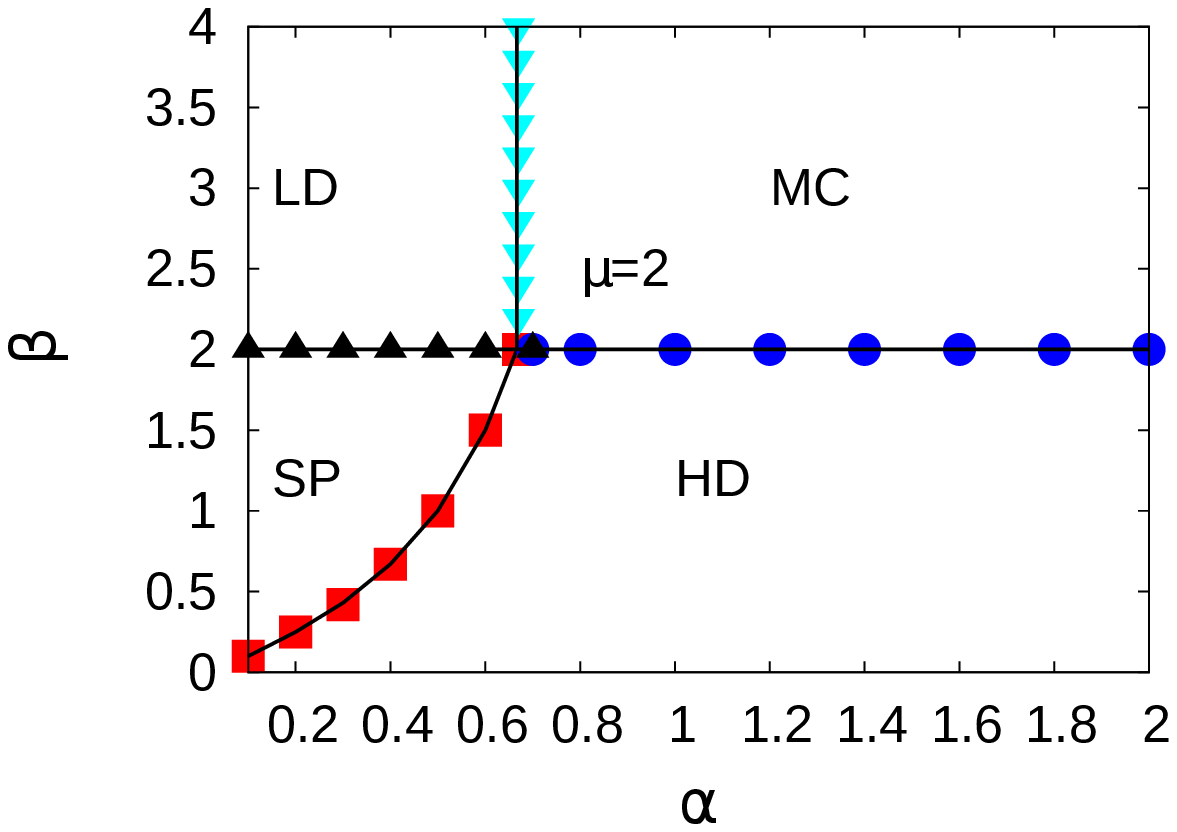}
 \caption{Phase diagram for $\mu=2$ in the $\alpha$ and $\beta$ plane. Continuous lines represent the MFT predictions (see text); the discrete points are the corresponding MCS results, which agree well with the MFT predictions. }\label{phase2}
\end{figure}
\begin{figure}
 \includegraphics[width=9cm]{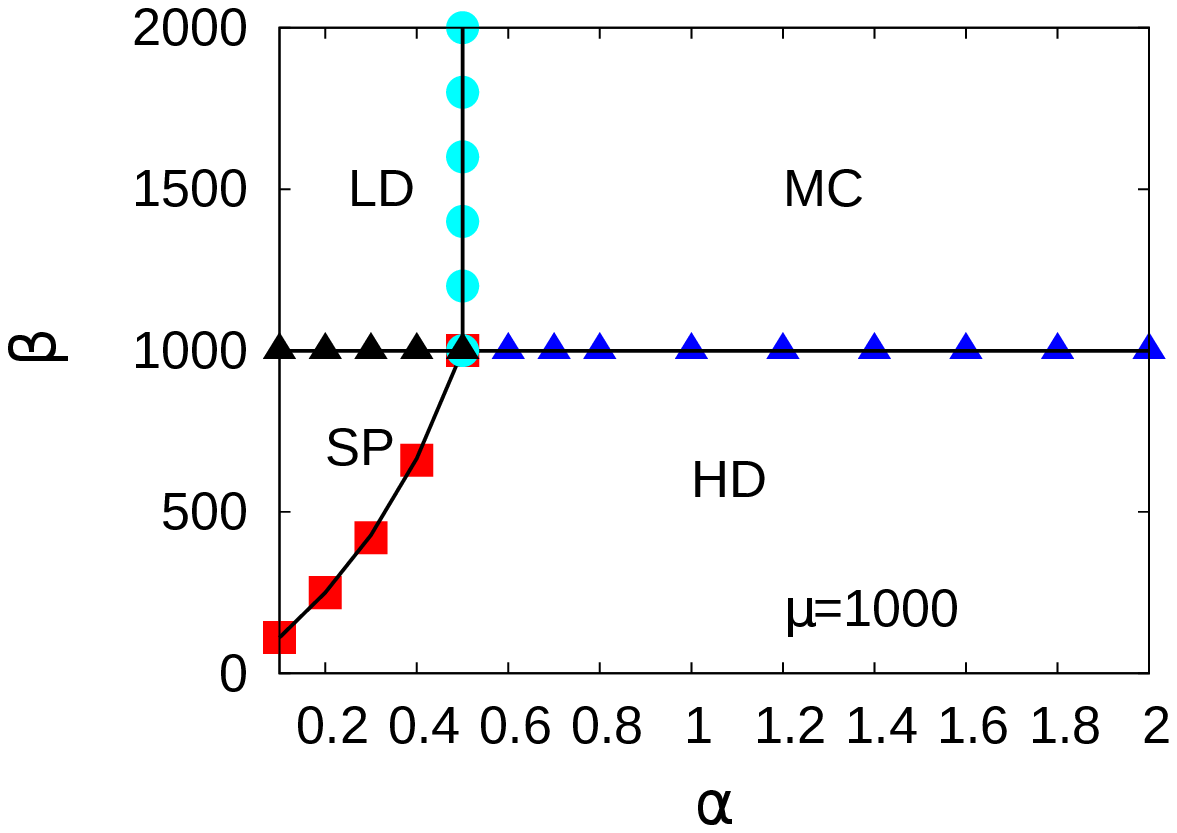}
 \caption{Phase diagram for $\mu=1000$ in the $\alpha$ and $\beta$ plane. Continuous lines represent the MFT predictions (see text); the discrete points are the corresponding MCS results, which agree well with the MFT predictions. }\label{phase3}
\end{figure}
\begin{figure}
 \includegraphics[width=9.3cm]{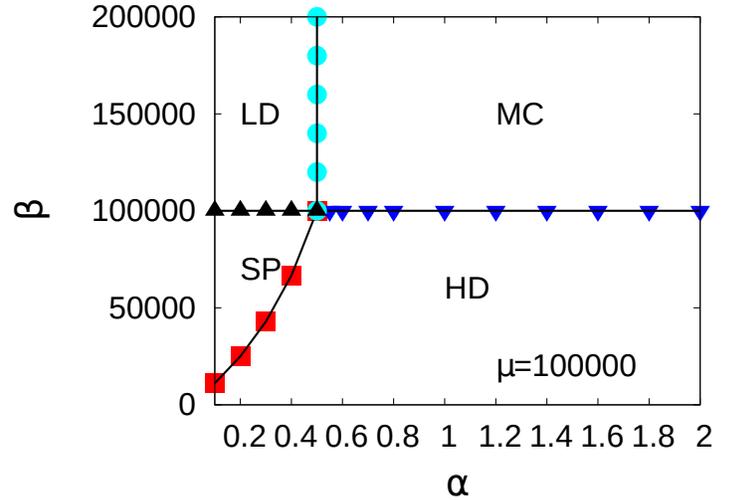}
 \caption{Phase diagram for $\mu=100000$ in the $\alpha$ and $\beta$ plane. Continuous lines represent the MFT predictions (see text); the discrete points are the corresponding MCS results, which agree well with the MFT predictions. }\label{phase5}
\end{figure}

Direct visual inspections of these phase diagrams reveal the following qualitative features. There are four phases LD, HD, MC and SP - in these phase diagrams. Further, all the phase boundaries meet at a single point. Then at $\mu=1$ all the phase boundaries are straight lines, whereas for all $\mu\neq 1$, except for the SP-HD boundary, all other boundaries are again straight lines. In contrast, the SP-HD boundary is a curved line for $\mu\neq 1$, whose shape appears to change as $\mu$ crosses unity.  {The phase diagrams in Fig.~\ref{phase2} ($\mu=2$), Fig.~\ref{phase3} ($\mu=1000$) and Fig.~\ref{phase5} ($\mu=100000$) have very similar forms; however, the scales of $\beta$-axis in these figures differ hugely.} We now quantitatively discuss these phase diagrams.

\section{Mean-field theory}\label{mft}

We now use MFT to construct the principles behind obtaining the phase diagrams and the associated density profiles. We recall that the instantaneous configuration of $T$ is described by a set of occupation numbers, that can take values 0 or 1, 
for each of the sites in $T$.
MFT entails neglecting spatial correlations and taking continuum limit with $\rho(x)\equiv \langle n(i)\rangle$, denoting the steady state 
densities in $T$; $x=i/L$ becomes quasi-continuous in the thermodynamic limit $L\rightarrow \infty$~\cite{tasep-mft}. Here $x$ 
starts from 0 at the entry end with $x=1$ at the exit end. In our MFT analysis, we study the phases of $T$ in terms of the well-known phases of a TASEP lane with open boundaries, 
delineated by the effective entry ($\alpha_{e}$) and exit ($\beta_{e}$) rates, respectively~\cite{hauke,parna-anjan}.  

\subsubsection{Low density phase}

\begin{figure}
\includegraphics[width=9cm]{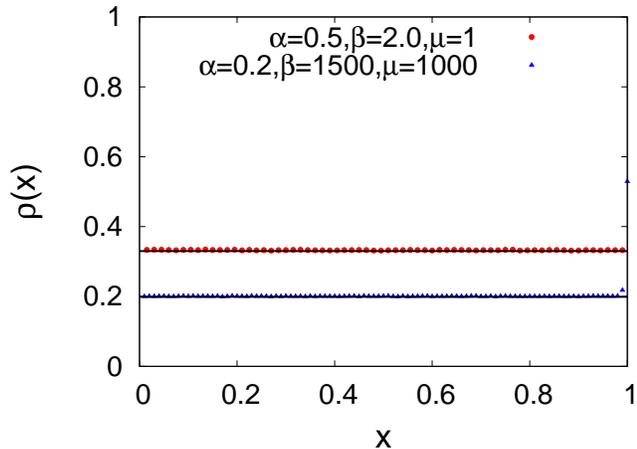}
 \caption{ Plots of $\rho(x)$ in the LD phase with different values for the parameters: $\mu=1000$, $\alpha=0.2$ and $\beta=1500$, and $\mu=1$, $\alpha=1/2$ and $\beta=2$.  Continuous lines represents the MFT predictions (see text), discrete points are the corresponding MCS results. Very good agreements between the MFT and MCS predictions are found. Notice that for $\mu=1000$, bulk density $\rho_{LD}$ is very close to $\alpha$, where as for $\mu$, it is substantially less than $\alpha$ (see text). Unsurprisingly, $\rho_{LD}$ is independent pf $\beta$.}\label{ld1}
\end{figure}

We begin with the LD phase. 
In the LD phase, we get by using (\ref{func})
\begin{equation}
\rho_{LD}=\alpha_{e} =\alpha \frac{N}{N_0}.
\end{equation}
Total particle number is given by
$N_0=N+L\alpha_e$. This implies for the total particle number 
\begin{equation}
N_0=N(1+\frac{\alpha L}{N_0}).
\end{equation}
This then gives
\begin{equation}
\rho_{LD}=\frac{\alpha}{1+\frac{\alpha}{\mu}},\label{ld-val}
\end{equation}
giving $\rho_{LD}<\alpha$, as expected.
See Fig.~\ref{ld1} for  representative plots of the steady state density profiles in the LD phase with different parameter values: $\mu=1000$, $\alpha=0.2$ and $\beta=1500$, and 
$\mu=1$, $\alpha=1/2$ and $\beta=2$.

\begin{figure}
 \includegraphics[width=9cm]{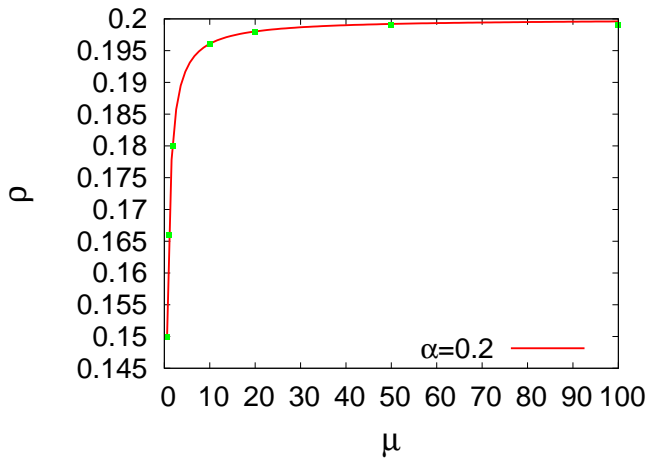}
 \caption{Plot of $\rho_{LD}$ versus $\mu$ for a given $\alpha=0.2$. The continuous black line represents the MFT result given by Eq.~(\ref{ld-val}); the points represent the MCS data. Clearly, $\rho_{LD}$ approaches $\alpha$ as $\mu$ grows (see text). Very good agreements between the MFT and MCS predictions are found. }\label{ld-plot}
\end{figure}

{ From Eq.~(\ref{ld-val}), we note that as $\mu$ grows, $\rho_{LD}$ approaches $\alpha$ as it would be for an open TASEP (for $\alpha<1/2$); see Fig.~\ref{ld-plot}.}

\subsubsection{High density phase}

\begin{figure}
 \includegraphics[width=9cm]{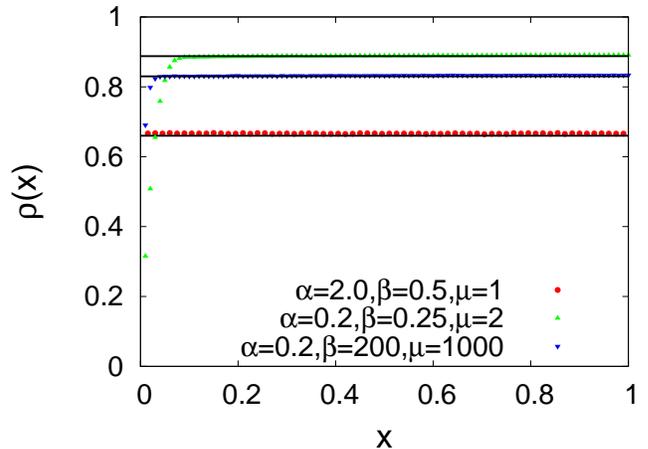}
 \caption{ Plots of $\rho(x)$ in the HD phase for $\mu=2$, $\alpha=0.2$ and $\beta=0.25$, and $\mu=1000$, $\alpha=0.2$ and $\beta=200$, and $\mu=1$, $\alpha=2$ and $\beta=1/2$.  Continuous  lines represent the MFT predictions (see text), discrete points are the corresponding MCS results. Very good agreements between the MFT and MCS predictions are found. Notice that $\rho_{HD}$ can be both smaller or larger than $1-\beta$ (see text). Unsurprisingly, $\rho_{HD}$ is independent of $\alpha$. }\label{hd1}
\end{figure}

Proceeding similarly for HD phase, we find
\begin{equation}
\rho_{HD}=1-\beta+\frac{\beta(1-\frac{1-\beta}{\mu})}{1+\frac{\beta}{\mu}}=\frac{\mu}{\beta+\mu}.\label{hd-val}
\end{equation}
Thus, $\rho_{HD}$ can be more or less than $1-\beta$, the HD phase bulk density in an open TASEP with an exit rate $\beta$.
See Fig.~\ref{hd1} with ($\mu=2$, $\alpha=0.2$ and $\beta=0.25$),  ($\mu=1000$, $\alpha=0.2$ and $\beta=200$) and  ($\mu=1000$, $\alpha=0.2$ and $\beta=200$).

\begin{figure}
 \includegraphics[width=9cm]{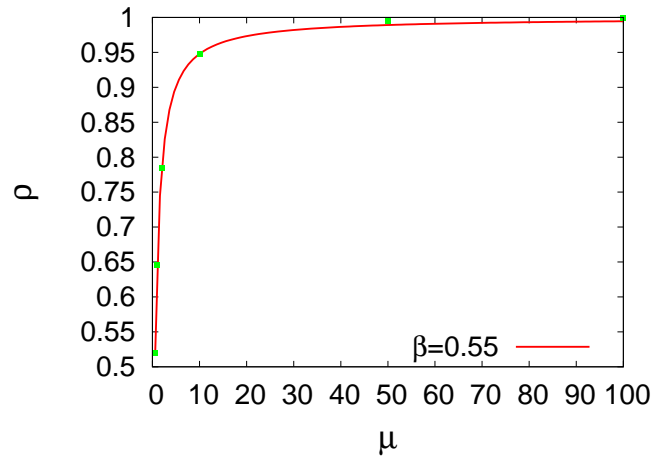}
 \caption{Plot of $\rho_{HD}$ versus $\mu$ for  given $\beta=0.55$. The continuous line represents  the MFT result given by Eq.~(\ref{hd-val}); the points represent the MCS data. Clearly, $\rho_{HD}$ approaches unity as $\mu$ becomes large (see text). Very good agreement between MFT and MCS results are found.}\label{hd-plot}
\end{figure}

 As revealed by Eq.~(\ref{hd-val}), $\rho_{HD}$ approaches {\em unity} (and hence independent of $\beta$) as $\mu$ becomes large; see Fig.~\ref{hd-plot}. This is in contrast with and marks a significantly departure from an open TASEP.

\subsubsection{Maximal current phase}

{ The MC phase is characterised by $\rho(x)=1/2$ in the bulk.} Hence, at the transition between LD to MC phase  $\rho_{LD}=\frac{1}{2}$. This gives the condition
\begin{equation}
\alpha=\frac{1}{2-\frac{1}{\mu}}.\label{ld-mc}
\end{equation}
Since $\alpha>0$ by definition, $\mu$ must be larger than 1/2 for the boundary (\ref{ld-mc}) to exist. In fact, there is {\em no} MC phase if $\mu < 1/2$, in agreement with our heuristic argument above. Notice that the boundary (\ref{ld-mc}) between the LD and MC phases is independent of $\beta$, and hence parallel to the $\beta$-axis in the $\alpha-\beta$ plane; see the phase diagrams in Fig.~\ref{phase1}, Fig.~\ref{phase4}, Fig.~\ref{phase2}, Fig.~\ref{phase3} and Fig.~\ref{phase5} 
as obtained from our MFT as 
well as our MCS simulation studies.

Similarly, at the transition between HD to MC phase, one has  $\rho_{HD}=\frac{1}{2}$ which in turn gives 
\begin{equation}
\beta=\mu.\label{hd-mc}
\end{equation}
 The boundary (\ref{hd-mc}) between the HD and MC phases is again a straight line in the $\alpha-\beta$ plane, in this case parallel to the $\alpha$-axis,  as can be seen in the phase diagrams in Fig.~\ref{phase1}, Fig.~\ref{phase4}, Fig.~\ref{phase2}, Fig.~\ref{phase3} and Fig.~\ref{phase5}.

\begin{figure}
 \includegraphics[width=9cm]{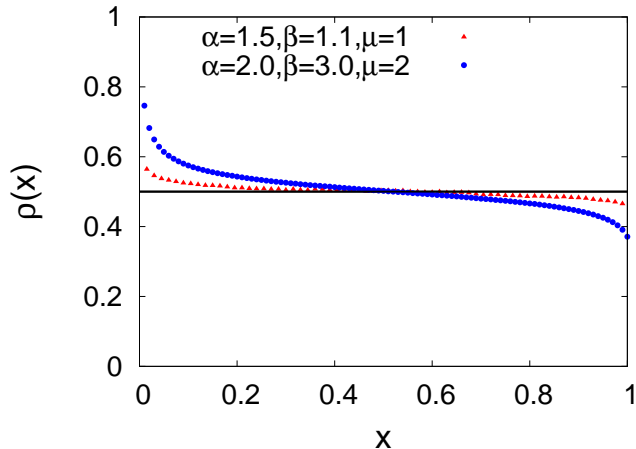}
 \caption{ Plots of $\rho(x)$ in the MC phase for $\mu=1$, $\alpha=1.5$ and $\beta=1.1$, and $\mu=2$, $\alpha=2.0$ and $\beta=3.0$. Continuous black lines represent the MFT predictions (see text), discrete points are the corresponding MCS results. Good agreements between the MFT and MCS predictions are found. Unsurprisingly, the bulk density (=1/2) is independent of $\alpha,\,\beta$; the deviations from 1/2 near the boundaries are the boundary layers that exist even in open TASEPs.}\label{mc-mu1}
\end{figure}

\subsubsection{Shock phase and domain walls}

The transition between the LD and the HD phases is marked by the condition
\begin{equation}
\rho_{LD} + \rho_{HD}=1, \label{dw-cond}
\end{equation}
which is equivalent to $\alpha_e=\beta_e$.
In open TASEPs, this transition is marked by a single 
delocalised domain wall (DDW) that spans the whole length of the TASEP. This is usually attributed to uncorrelated entry and 
exit events in an 
open TASEP. In the present model, there is a strict number conservation. This leads to a localised domain wall (LDW), as opposed to a DDW in an open 
TASEP~\cite{hauke,parna-anjan}. We discuss this in details below.

In order to characterise the SP  fully, we must obtain the domain wall height $\Delta$ and the domain wall position $x_w$ as functions of the model parameters.  From particle number conservation, we get 
\begin{equation}
                                    N_0=L\int_{0}^{1}\rho(x)\,dx+N.
                                   \end{equation}
This implies,
\begin{eqnarray}
 \mu&=&\int_{0}^{x_w}\,dx\,\alpha_{e}+\int_{x_w}^{1}(1-\beta_{e})dx+N/L\nonumber\\
 &=&\alpha_{e}x_w+(1-\beta_{e})(1-x_w)+N/L\nonumber\\
 &=&\frac{N\alpha}{N_0}x_w+1-x_w-\beta(1-\frac{N}{N_0})\nonumber \\&+&\beta(1-\frac{N}{N_0})x_w+\frac{N}{L}.\label{mu-eq}
 \end{eqnarray}
 At the LD-HD coexistence, we have
 \begin{equation}
  \alpha_e=\beta_e,
  \end{equation}
  This gives
  \begin{eqnarray}
 \alpha \frac{N}{N_0}&=&\beta(1-\frac{N}{N_0}),\nonumber\\
 \implies(\alpha+\beta)\frac{N}{N_0}&=&\beta,\nonumber\\
 \implies\frac{N}{N_0}&=&\frac{\beta}{\alpha+\beta}.\label{shock-N}
 \end{eqnarray}
 Thus, $N/N_0$ is {\em independent} of $\mu$.
 Hence,
 \begin{equation}
  \alpha_e = \frac{\alpha\beta}{\alpha+\beta}=\beta_e.\label{alpha-beta-e}
 \end{equation}

Using the above equation in (\ref{mu-eq}) for $\mu$ we write
\begin{eqnarray}
 \mu&=&x_w\alpha\frac{\beta}{\alpha+\beta}+1-x_w-\beta+\beta\frac{\beta}{\alpha+\beta}\nonumber \\&+&\beta x_w-x_w\beta\frac{\beta}{\alpha+\beta}+\mu\frac{\beta}
 {\alpha+\beta}.
\end{eqnarray}
Simplifying the above equation we get
\begin{equation}
 x_w=\frac{\mu\alpha-\alpha-\beta+\alpha\beta}{2\alpha\beta-\alpha-\beta}.\label{xw-gen}
\end{equation}
This gives the position of DW. Note that for fixed $\alpha,\,\beta$, $x_w$ changes continuously with $\mu$. For the SP phase to exist, $0<x_w<1$. For HD to SP transition $x_w=0$, which in turn implies,
\begin{equation}
 \beta=\frac{\alpha(\mu-1)}{1-\alpha}.\label{hd-sp}
\end{equation}
For LD to SP transition $x_w=1$ which implies,
\begin{equation}
 \beta=\mu.\label{ld-sp}
\end{equation}
In the $\alpha-\beta$ plane, (\ref{hd-sp}) is clearly not a straight line in general, where (\ref{ld-sp}) is a straight line. See Fig.~\ref{phase1}, 
Fig.~\ref{phase2}, Fig.~\ref{phase3} and Fig.\ref{phase5}. For $\mu=1$, the phase boundary is however a straight line; see Fig.~\ref{phase4}. { The SP phase is thus confined between 
the lines (\ref{ld-sp}) and (\ref{hd-sp}), and hence covers {\em a region} in the $\alpha-\beta$ plane.}

\begin{figure}
 \includegraphics[width=9cm]{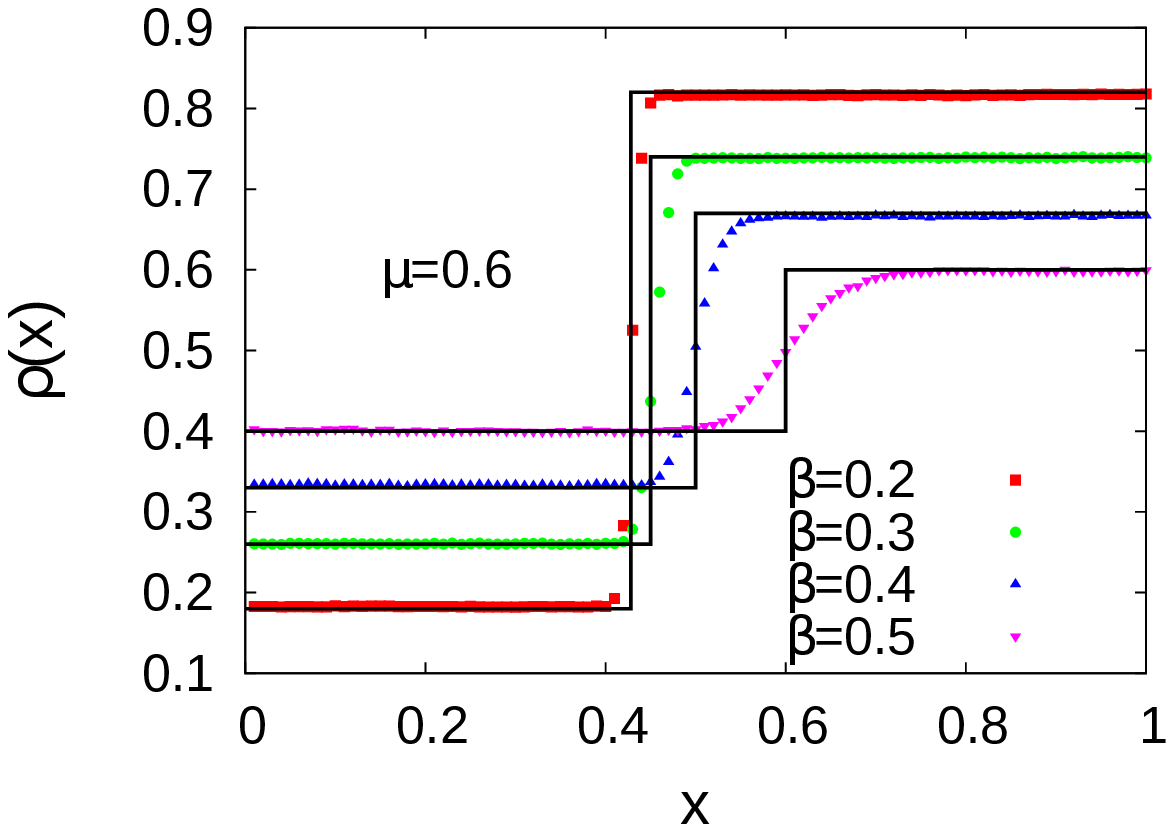}
 \caption{LDW in $T$ for $\mu=0.6$, $\alpha=4.0$ and different values of $\beta$. Continuous black line represents the MFT prediction (see text),  points with in various colours are the corresponding MCS results.}\label{dw-mu06}
\end{figure}

\begin{figure}
 \includegraphics[width=9cm]{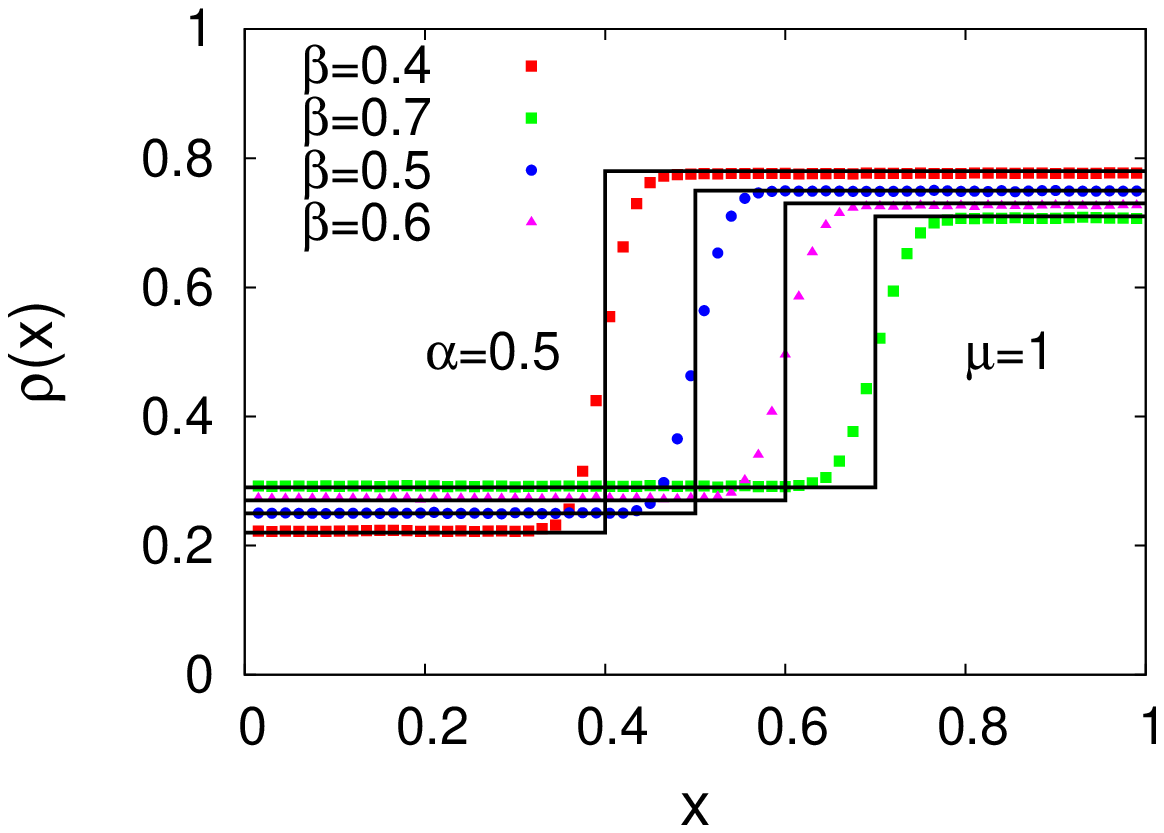}
 \caption{LDW in $T$ for $\mu=1$, $\alpha=0.5$ and different values of $\beta$. Continuous black line represents the MFT prediction (see text),  points in various colours are the corresponding MCS results.}\label{dw-mu1}
\end{figure}

\begin{figure}
 \includegraphics[width=9cm]{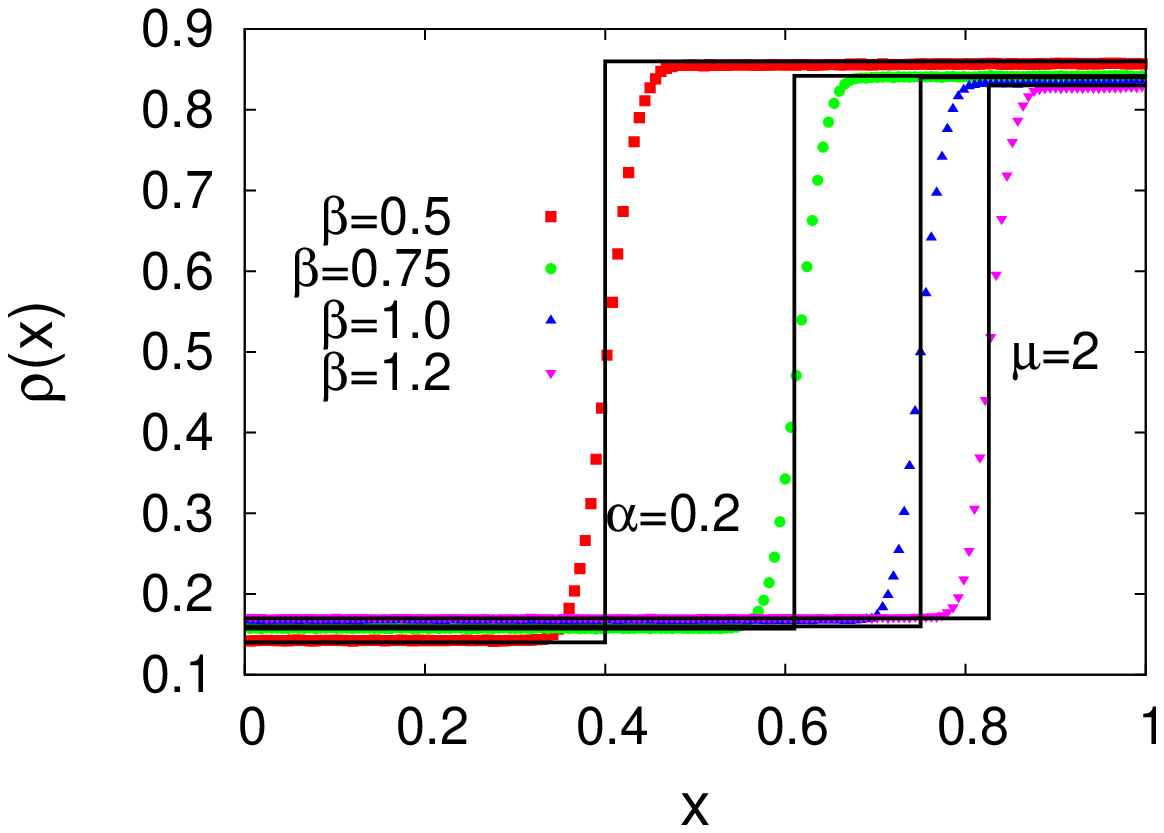}
 \caption{LDW in $T$ for $\mu=2$, $\alpha=0.2$ and different values of $\beta$. Continuous black line represents the MFT prediction (see text),  points in various colours are the corresponding MCS results.}\label{dw-mu2}
\end{figure}

\begin{figure}
 \includegraphics[width=9cm]{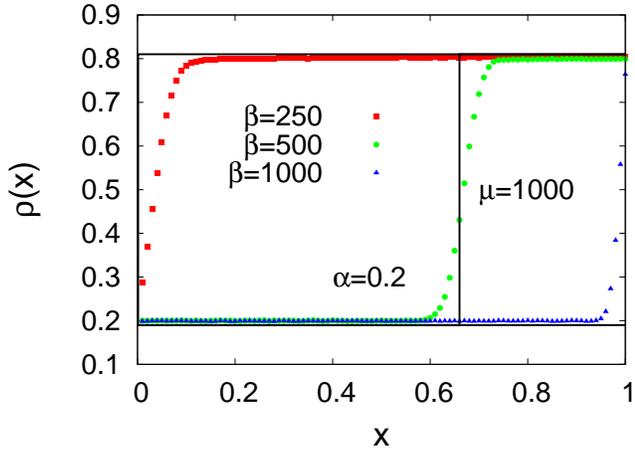}
 \caption{LDW in $T$ for $\mu=1000$, $\alpha=0.2$ and different values of $\beta$. Continuous black line represents the MFT prediction (see text), points in various colours are the corresponding MCS results.}\label{dw-mu1000}
\end{figure}

We now find the height of the domain wall. Noting that at the entry 
side, $\rho(x)$ has a mean value $\rho_{LD}=\alpha_e$, where as on the exit side, it $\rho_{HD}=1-\beta_e=1-\alpha_e$, since 
$\alpha_e=\beta_e$ for a domain wall to exist, we find the domain wall height $\Delta$ as
\begin{equation}
 \Delta = \rho_{HD}-\rho_{LD}=1-2\alpha_e = 1-2\frac{\alpha\beta}{\alpha+\beta}.\label{delta-dw}
\end{equation}
See Fig.~\ref{dw-mu06}, Fig.~\ref{dw-mu1}, Fig.~\ref{dw-mu2} and Fig.~\ref{dw-mu1000} for representative 
plots of LDW for various values of the model parameters. 

{
It is clear from these figures that the LDW height $\Delta$ is a function of $\alpha,\,\beta$, but not depend upon $\mu$ [{\em cf.} Eq.~(\ref{delta-dw}); see also Fig.~\ref{delta-dw-fig}]. In contrast, its position $x_w$ depends  on all three of $\alpha,\,\beta$ and $\mu$ [{\em cf.} Eq.~(\ref{xw-gen})]. In fact from Eq.~(\ref{xw-gen}), we note that if $\alpha=1/2$, $x_w=\beta-(\mu-1)$, a straight line as a function of $\beta$, but is a generic nonlinear function of $\beta$ for all other values of $\alpha$; $x_w$ retains a linear dependence on $\mu$ for all $\alpha,\,\beta$ within SP.

In Fig.~\ref{xw-plot} and Fig.~\ref{delta-plot}, we have shown the dependence of $x_w$ and $\Delta$ on $\beta$ for given $\alpha,\,\mu$, as obtained from our MFT and MCS studies. 
\begin{figure}[htb]
 \includegraphics[width=9cm]{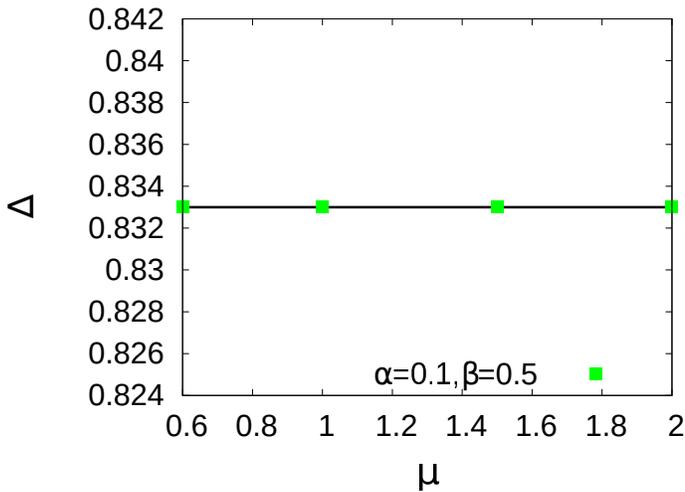}
 \caption{Plot of $\Delta$ versus $\mu$ for $\alpha=0.1,\,\beta=0.5$. The continuous line represents the MFT result for $\Delta$ given by Eq.~(\ref{delta-dw}); the discrete points are the corresponding MCS results. Very good agreements between MFT and MCS results can be seen. Clearly, $\Delta$ does not depend on $\mu$ within SP (see text).} \label{delta-dw-fig}
\end{figure}

\begin{figure}
 \includegraphics[width=8cm]{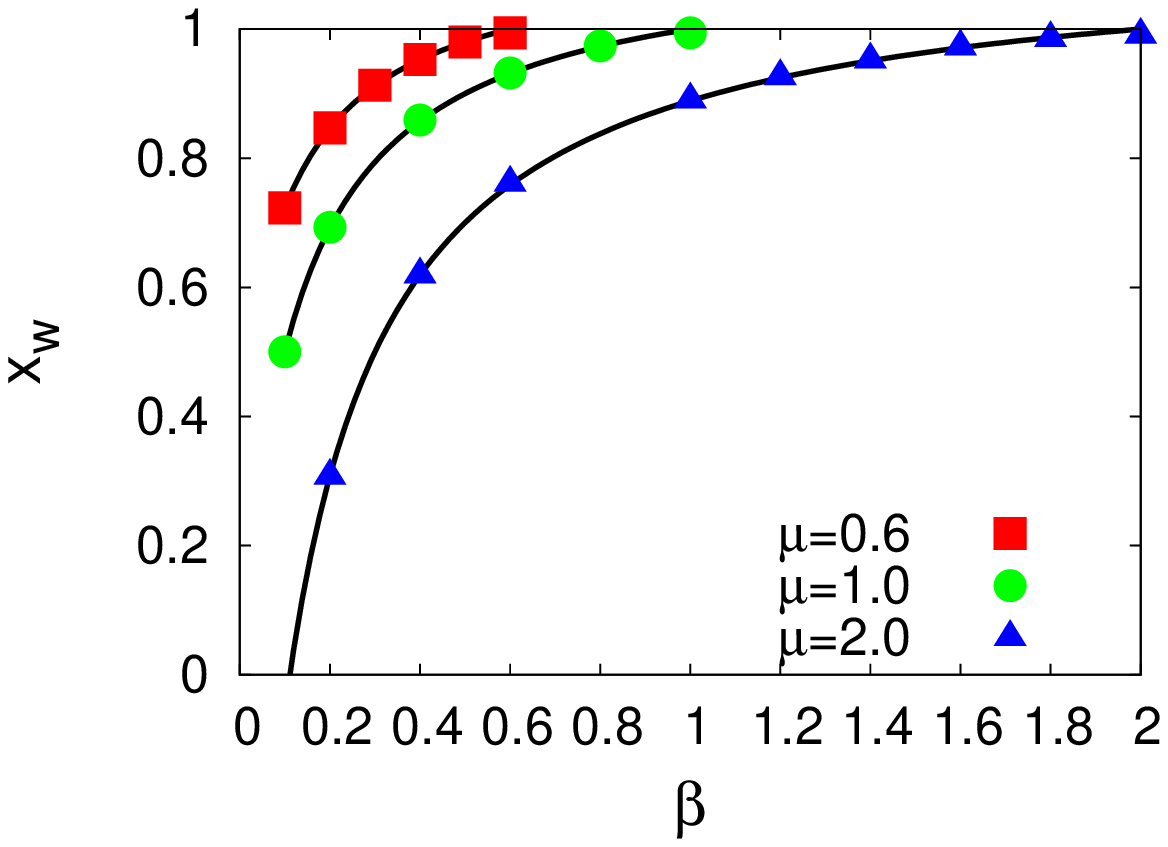}\\
 \includegraphics[width=8cm]{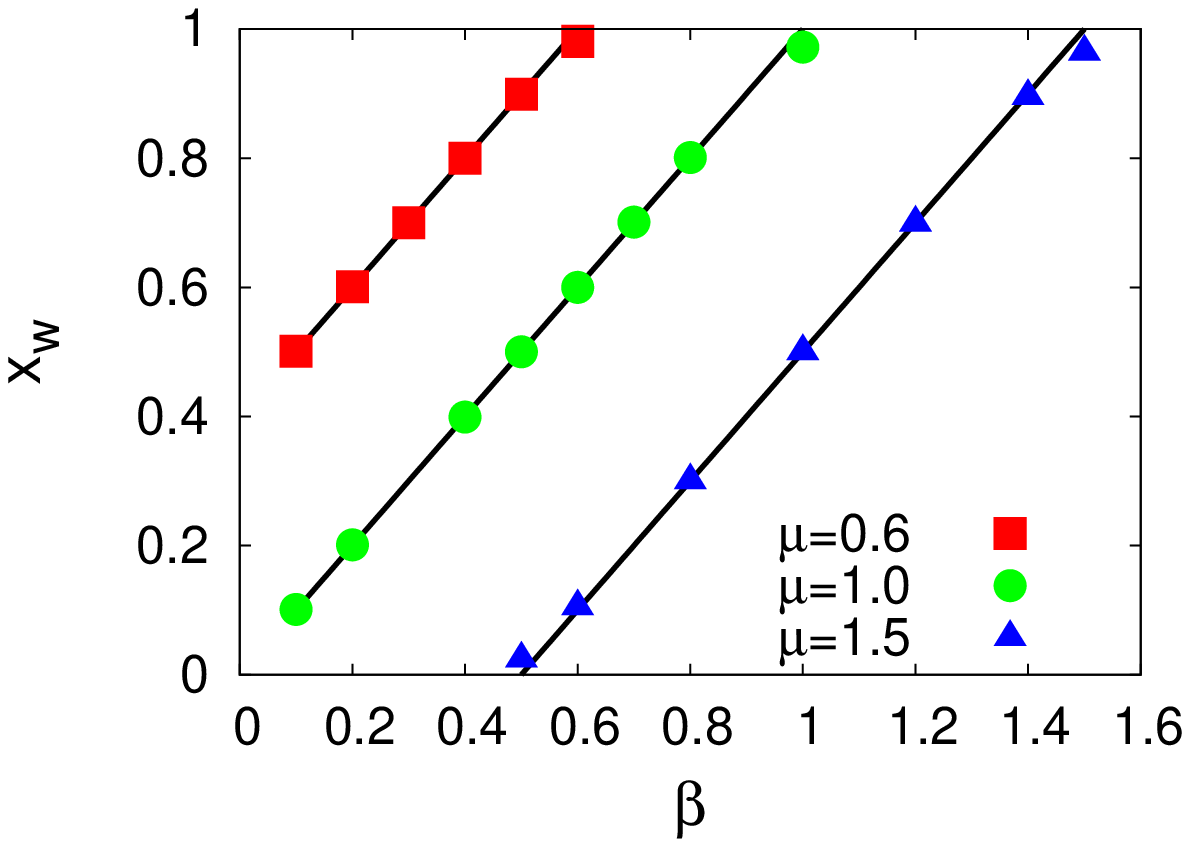}
 \caption{Plot of $x_w$ versus $\beta$ for (top) $\alpha=0.1$ and (bottom) $\alpha=0.5$. The continuous line represents the MFT result for $x_w$ given 
 by Eq.~(\ref{xw-gen}); the discrete points are the corresponding MCS results. For $\alpha=0.1$, the dependence of $x_w$ on $\beta$ is not linear, where as for $\alpha=0.5$, it depends linearly (see text). Very good agreements between MFT and MCS results can be seen.}\label{xw-plot}
\end{figure}
\begin{figure}
 \includegraphics[width=8cm]{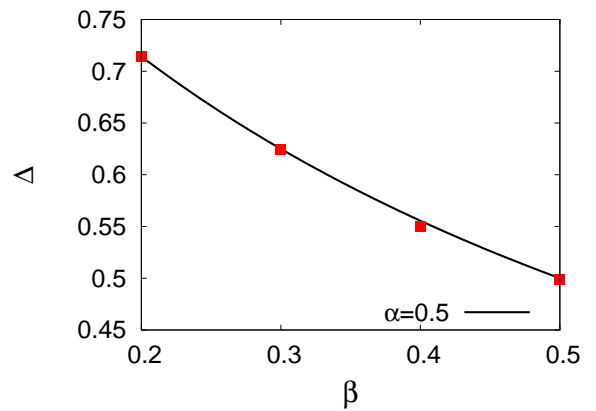}
 \caption{Plot of $\Delta$ versus $\beta$ for $\alpha=0.5$ and $\mu=0.6$ and $\mu=1$. The continuous line represents the MFT result for $\Delta$ given by 
 Eq.~(\ref{delta-dw}); the points represent the corresponding MCS results, which overlap. Very good agreements between the MFT and MCS results can be seen. }\label{delta-plot}
\end{figure}

The reason that $\Delta$ is independent of $\mu$ lies in Eq.~(\ref{shock-N}), which shows that the reservoir occupation $N$ {\em does not} depend upon $\mu$ in the SP phase. Thus, as $\mu$ is varied within the SP phase, $N$ does not change; as a result $\alpha_e$ and $\beta_e$ do not change which in turn leaves $\Delta$ unchanged. As $\mu$ changes $N_0$ however changes. These extra or deficit particles are adjusted by changing the position $x_w$ of the DW~\cite{foot2}.}


\subsubsection{Phase boundaries meet a common point}

\begin{figure}
 \includegraphics[width=8cm]{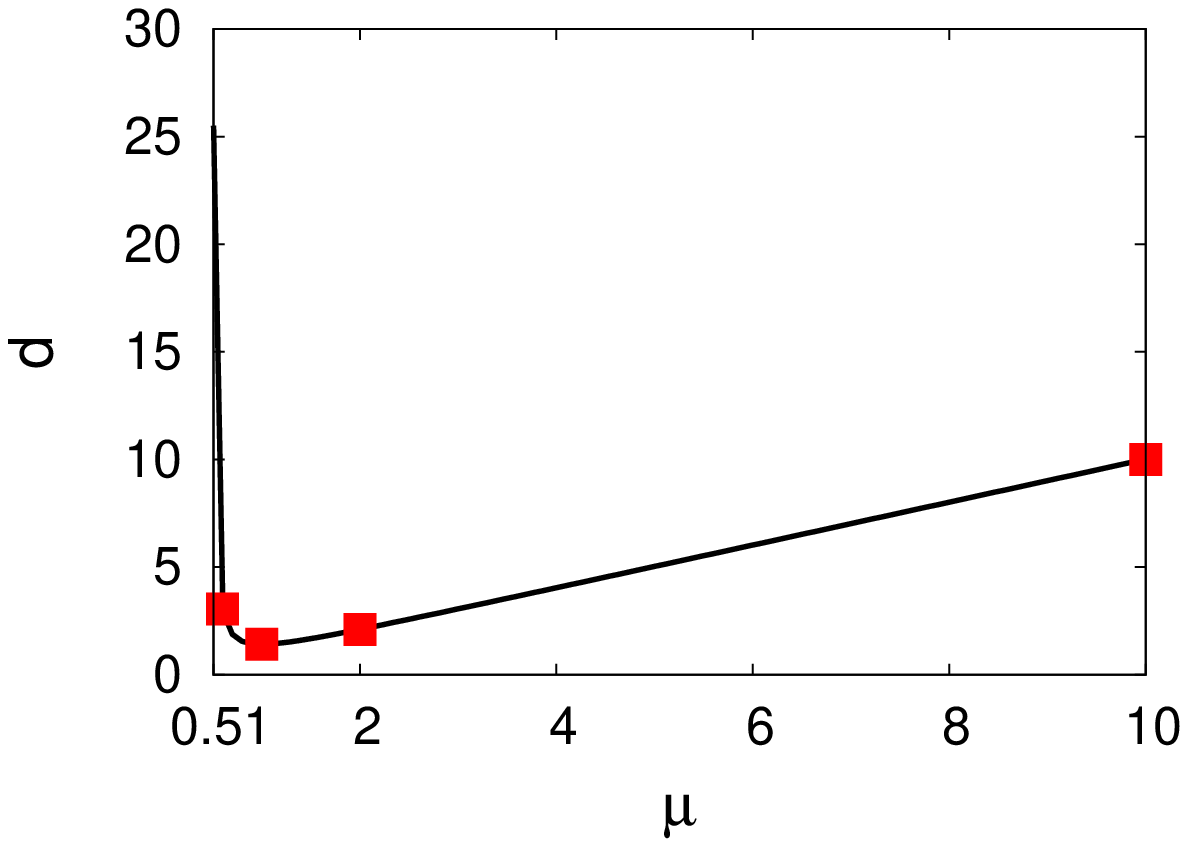}
 \caption{Plot of $d(\mu)$ versus $\mu$ in the SP phase. The continuous line represents the MFT result given by Eq.~(\ref{dmu-eq}); the discrete points are the corresponding MCS results. Clearly $d$ diverges for when either $\mu\rightarrow 1/2$, or $\mu\rightarrow \infty$, showing the absence of the four phase coexistence point as these two limits are approached (see text). Very good agreement between MFT and MCS results are found.}\label{d-mu}
\end{figure}

All the four phase boundaries meet at a common point $(\alpha_c,\,\beta_c)=(\mu/(2\mu -1),\,\mu)$, which is a {\em multicritical point}, as we explain below in the next Section. It is, however, useful to consider the ``distance'' $d$ between the origin $(0,\,0)$ and $(\alpha_c,\,\beta_c)$ as a function of $\mu$: We find
\begin{equation}
 d=\sqrt{\frac{\mu^2}{(2\mu-1)^2} +\mu^2}.\label{dmu-eq}
\end{equation}
Thus, $d$ diverges when $\mu\rightarrow 1/2$ from above, or when $\mu\rightarrow\infty$;
see Fig.~\ref{d-mu} for a plot of $d(\mu)$ versus $\mu$. Both MFT and MCS results are shown, which agree well with each other.

In the limit of infinite capacity, i.e.,
$\mu\rightarrow\infty$, $(\alpha_c,\,\beta_c)\rightarrow (1/2,\,\infty)$. { Since $\beta_c\rightarrow \infty$, the latter phase essentially is confined to the $\beta$-axis (see Section~\ref{infin-phase}) below.} On the other hand, as $\mu\rightarrow 1/2_+$, 
$(\alpha_c,\,\beta_c)\rightarrow (\infty,\,1/2)$. Thus, for $\beta<1/2$ and all $\alpha$, the TASEP shows an LDW, where as for 
$\beta>1/2$ and 
all $\alpha$, the system is in the LD phase. For $\mu <1/2$, $\alpha_c$ becomes negative, which is unphysical. As heuristically argued above, this implies that for 
$\mu <1/2$, 
the system can only be in the LD phase or SP, as there are not enough particles for HD or MC phases. This can be understood easily. Since HD or 
MC phase would require at least $\mu=1/2$ or more (assuming all particles are in $T$, leaving the reservoir empty), with $\mu<1/2$, there are 
enough particles just for 
LD and SP only.

The nature of the HD-SP boundary changes as $\mu$ crosses unity, as is evident from Eq.~(\ref{hd-sp}). For $\mu <1$, $\alpha >1$ necessarily. 
On the boundary, as $\alpha\, (>1)$ grows, $\beta$ decreases. In fact, in the limit $\alpha\rightarrow \infty$, $\beta\rightarrow 1-\mu$. In contrast 
for $\mu>1$, 
$\alpha<1$ at the boundary since $\beta$ cannot be negative. As $\alpha (<1)$ reduces, $\beta$ also reduces. This explains the difference between 
the shape of the HD-SP boundary in Fig.~\ref{phase1} ($\mu=0.6<1$) vis-a-vis in the other phase diagrams in Fig.~\ref{phase4}, Fig.~\ref{phase2}, Fig.~\ref{phase3} and Fig.~\ref{phase5} with $\mu>1$. The phase diagram in Fig.~\ref{phase4} for $\mu=1$ merits separate attention. At $\mu=1$, (\ref{xw-gen}) gives $\alpha=1$ as the boundary between the HD and SP (since $\beta \neq 0$). On the other hand, (\ref{ld-sp}) gives $\beta=1$ as the boundary between LD and SP for $\mu=1$. With $\mu=1$, all the four phases meet at $(1,\,1)$ in the $\alpha-\beta$ plane. This immediately gives the phase diagram in
Fig.~(\ref{phase4}).

\subsubsection{Phases for infinite resources}\label{infin-phase}

By using the logic of the MFT constructed above, we can now infer the admissible phases for $\mu\rightarrow \infty$. First of all, notice that at $\mu$ rises, the region of the phase space spanned by the HD phase rises, a feature that is evident from the phase diagrams presented above. It is thus reasonable to expect that with $\mu\rightarrow\infty$, the TASEP will be typically in the HD phase, as we now argue.  When $\mu\rightarrow\infty$, the reservoir occupation $N$ must also approach infinity. Due to the crowding effect, $\beta_e\rightarrow 0$ and $\rho_{HD}\rightarrow 1$ for any finite $\alpha$; see Eq.~(\ref{hd-val}). Thus the TASEP channel should be nearly filled, with the HD phase being the only possibly phase for any finite $\alpha$ and $\beta$. No other phase is to be observed, including no possibility of any domain wall for any finite $\alpha$ and $\beta$. This makes it significantly different from an open TASEP which can be in LD, HD or MC phases. That only the HD phase is possible can be seen 
from the fact that the 
multicritical point $(\
\alpha_c,\,\beta_c)$ moves to $(1/2,\infty)$ for $\mu\rightarrow \infty$. Thus the boundaries between the SP and LD phases and HD and 
MC phases all move to $\beta =\infty$. Further, the slope of the boundary between the SP and HD phases, as given by Eq.~(\ref{hd-sp}), diverges as $\mu\rightarrow \infty$, indicating that the shock phase essentially gets concentrated on the $\beta$-axis, leaving the entire phase diagram to be spanned by the HD phase. We can further argue that for $\mu\rightarrow 0$, there can only be the LD phase, for 
with $\mu\rightarrow 0$, there are too few particles to have any phase other than the LD phase.


\section{Phase transitions in the model}\label{phase-trns}

Phase diagrams in Fig.~\ref{phase1}, Fig.~\ref{phase4}, Fig.~\ref{phase2}, Fig.~\ref{phase3} and Fig.~\ref{phase5} all have different phases separated by sharp phase boundaries. We now discuss the 
nature of the transitions across these phase boundaries. In an open TASEP, the transitions between the LD and HD phases are accompanied by a sudden 
jump in the bulk density in the TASEP, which indicates a first order transition with the steady state bulk density acting as the order parameter. The corresponding phase boundary 
is characterised by a single DDW. In the same vein, the transitions between the LD or HD and MC phases are second order transitions, with the density changing smoothly at the phase transition. In the phase diagram of an open TASEP, three phase boundaries - two second order (LD-MC and HD-MC) 
and one first order (LD-HD) boundaries - meet at a multicritical point. In contrast, the phase diagrams for the present model generically all have four phase boundaries, 
one each for the transition between LD-MC, LD-SP, MC-HD and SP-HD phases. Notice that, unlike for an open TASEP, there is {\em no} phase boundary that acts as the boundary between the LD and HD phases. In other words, as one moves in the parameter space, one cannot directly move from the LD to the HD phases and vice versa. Again taking the steady state bulk density as the order parameter, we note that the density 
changes {\em smoothly} across all the four phase boundaries. Thus, all the transitions and the associated phase boundaries represent  {\em second order} transitions. 
All these four second order lines meet at a multicritical point $(\alpha_c,\,\beta_c)$. This feature of the phase diagram is similar to that in the model studied in Refs.~\cite{hauke,brackley}, and in one of the models studied in Refs.~\cite{parna-anjan}. Lastly, for $\mu<1/2$, there are only two phases - LD and SP - possible, and the multicritical point naturally ceases to exist for $\mu<1/2$.

\section{Nature of the domain walls}\label{dw}

As a consequence of the strict particle number conservation, MFT gives the precise location of the DW [see Eq.~(\ref{xw-gen})], implying an LDW; see Fig.~\ref{dw-mu06}, Fig.~\ref{dw-mu1}, Fig.~\ref{dw-mu2} and Fig.~\ref{dw-mu1000} for representative 
plots of LDW for various values of the model parameters. It is clear that the domain wall is {\em always} sharply pinned for all the choices 
of $\mu$. This is in contrast to a DDW in an open TASEP that has no particle number conservation in it. Still, the LDW does fluctuate about its mean position, as the particle number conservation applies on the whole system, and not on the TASEP segment, leaving the particle content 
in the TASEP to fluctuate (subject to maintaining the overall conservation). Indeed, it was shown recently~\cite{parna-anjan} that in a 
ring geometry consisting of a TASEP segment and a diffusive segment, in certain limits of the model parameters when the variance of the TASEP particle content scales with the size of the 
TASEP or the relative particle content of the diffusive segment, the LDW gets depinned and takes a form identical to that of a DDW in an open TASEP.   This was explained in terms of the attendant diverging particle fluctuations in the TASEP segment necessary for the formation of a DDW~\cite{parna-anjan}. It was further demonstrated that this transition from an LDW to a DDW is a smooth crossover with the LDW fluctuations increase gradually with the extent of particle number fluctuations in the TASEP segment, which in turn grows with the size of the 
diffusive segment, or the relative particle content of the diffusive segment. Analogous to the studies in Ref.~\cite{parna-anjan}, it is expected that with rising $\mu$, i.e., with increasing total particle numbers in the system, the 
LDW fluctuations should increase leading to its eventual delocalisation and formation of a DDW.  Surprisingly and unexpectedly, 
no such tendency towards eventual delocalisation of the LDW has been observed in the MCS studies of the present model. We now explain this unexpected lack of delocalisation by systematically studying its fluctuations.

We now analyse  the domain wall fluctuations and closely follow Refs.~\cite{tobias,tirtha-prr} in this Section. We consider a domain wall with a  position at $x_w$, that fluctuates in time. With $\Delta$ as the height of the DW, increasing the number of particles by one implies shifting the instantaneous DW by $\delta x_w=-1/(L\Delta)$. Similarly, a particle exiting the TASEP means shifting the DW by $\delta x_w = +1/(L\Delta)$. Let $P(x_w,t)$ be the probability 
of finding the DW at $x_w$ at time $t$. Then, following Refs.~\cite{tobias,tirtha-prr}, we find that $P$ satisfies the Fokker-Planck equation
\begin{equation}
 \frac{\partial P}{\partial t} = D\frac{\partial^2}{\partial y^2}P,
\end{equation}
where $y\equiv \delta x_w$ and $D$ is a diffusion constant given by
\begin{equation}
 D=\frac{1}{2}[\alpha_e(1-\alpha_e) + \beta_e(1-\beta_e)].
\end{equation}
As shown above, in the limit $\mu\rightarrow\infty$, $1-\beta_e=\rho_{HD}\rightarrow 1$. This implies $\beta_e\rightarrow 0$. Since 
$\alpha_e=\beta_e$ for a DW, we must have $\alpha_e\rightarrow 0$ as well. Thus, $D\rightarrow 0$ for $\mu\rightarrow \infty$. Therefore, the corresponding time scale of fluctuations $1/D$ diverges for a fixed  $L$. Thus, even for a finite $L$, the typical time required for the DW to traverse the whole of TASEP of size $L$ diverges, making any delocalisation essentially unobservable.  

\section{Summary and outlook}\label{sum}

In this article, we have explored the ``crowding effect'' of the reservoir on the attached TASEP within a simple model. To this end, we have studied the nonequilibrium steady states of a TASEP connected at both its ends to a reservoir  without any internal dynamics. Both the effective entry and exit rates to and from the TASEP, respectively, depend on the reservoir occupation number, that is in turn dynamically determined in this model. The presence of dynamically-controlled effective rates mean  a rising reservoir occupation can hinder flow of particles from TASEP back to the reservoir, but facilitates particle flow into TASEP and vice versa - a property that we have named ``crowding effect'' of the reservoir. The reservoir can accommodate an unlimited number of particles.  We have focused on how  the total particle number conservation conspires with the crowding effect to ultimately control the steady state density profiles and the phase diagram of the TASEP. This model generically shows a static or pinned 
domain wall or a single LDW, unlike for an open TASEP. Furthermore, the {\em shock phase}, where such an LDW is expected to be 
observed, is no longer a single line as for an open TASEP, rather covers a {\em region} in the parameter space spanned by the two entry and exit rate parameters. As a result, all the transitions in this model are {\em second order} in nature. While it is na\"ively expected that in the limit of infinite density, this model should reduce to an open TASEP, for in that limit, the effects of particle number conservation should vanish, we show that it does not happen: the large density limit of this model is {\em distinct} from an open TASEP, as it still allows only an LDW and not a DDW. This is primarily a consequence of the crowding effect, as argued here, which is never present in an open TASEP. { We have used analytical MFT and MCS studies for our work, and find very good agreement between the MFT and MCS results, that lends credence to our MFT.}

Our model may be generalised for future studies in several ways. For instance, by changing the precise dependence of $\alpha_e$ and $\beta_e$ on the reservoir occupation, one could study the sensitivity of the phase diagram on these dependences. It would be interesting to introduce diffusive exchanges or Langmuir kinetics between the reservoir and the bulk of the TASEP that does not violate the global conservation of particles, but breaks it locally in the bulk of the TASEP, and see whether the phase diagram can be changed significantly, or new phases can emerge. We have assumed the reservoir to be a point, devoid of any spatial extent and internal dynamics. While this makes the ensuing calculations simple and analytically tractable, this makes it an idealisation of more complex situations where internal reservoir dynamics is generically expected. This may be incorporated by using and suitably modifying some of the models studied in Ref.~\cite{parna-anjan}. Lastly, multiple TASEP lanes and multiple species 
of 
particles along with reactions 
between them would be an interesting future study that couples driven reactions with overall particle number conservation and crowding effects.


\begin{thebibliography}{99}
\bibitem{foot2} { That the domain wall $\Delta$ is independent of $\mu$ may also be understood heuristically as follows. In the SP phase, as $\mu$ changes keeping $\alpha$ and $\beta$ unchanged, the excess or deficit particles may be accommodated by either changing (i) the reservoir occupation $N$, or (ii) the domain wall height $\Delta$, or (iii) the domain wall position $x_w$, or all or some  of them together. Possibilities (i) and (ii) would have to happen together, if at all, since $\alpha_e,\,\beta_e$ (which control the height) are related to the reservoir occupation. Hence, a change in $N$  would then automatically imply changes in $\alpha_e$ [through Eq.~(\ref{ld-val})] and $\beta_e$ [through Eq.~(\ref{hd-val})]. Therefore, if these two possibilities were to be adopted by the system, and if there was a domain wall at some $(\alpha,\,\beta)$, there would not be any domain wall now when $\mu$ is changed to $\mu'$, as $\alpha_e(\mu')\neq \beta_e(\mu')$ now. Thus a domain wall could exist for a given 
$(\alpha,\,\beta)$ only for a particular value of $\mu$; a small change in $\mu$ to $\mu'$ would result into the system moving to the LD (if $\alpha_e(\mu')<\beta_e(\mu')$ or HD (if $\alpha_e(\mu')>\beta_e(\mu')$) phases completely. This necessarily leads to a rise in the steady state current. On the other hand, with possibility (iii) all the excess or deficit particles are accommodated by shifting the domain wall position $x_w$ keeping the height $\Delta$ intact (and also keeping the reservoir occupation $N$ intact). This ensures that the current is unchanged, independent of $\mu$ within the SP phase. Thus, within a minimum steady state current principle, possibility (iii) is the acceptable solution.}
\end{thebibliography}
\end{document}